%% file: master.tex
\documentclass[9pt,english,twocolumn,letter]{article}

\include{template/preamble}

\begin{document}
\author{
    Baptiste Wicht (EIA-FR)\\
    \texttt{baptiste.wicht@gmail.com}
    \and
    Roberto A. Vitillo\\
    \texttt{ra.vitillo@gmail.com}
    \and
    Dehao Chen (Google)\\
    \texttt{dehao@google.com}
    \and
    David Levinthal (Google)\\
    \texttt{levinth@google.com}
}
\title{\bf Hardware Counted Profile-Guided Optimization}
\date{}

\maketitle

\begin{abstract}

Profile-Guided Optimization (PGO) is an excellent means to improve the performance of a compiled program. Indeed, the execution path data it provides helps the compiler to generate better code and better cacheline packing.

At the time of this writing, compilers only support instrumentation-based PGO. This proved effective for optimizing programs. However, few projects use it, due to its complicated dual-compilation model and its high overhead. Our solution of sampling Hardware Performance Counters overcome these drawbacks. In this paper, we propose a PGO solution for GCC by sampling Last Branch Record (LBR) events and using debug symbols to recreate source locations of binary instructions.

By using LBR-Sampling, the generated profiles are very accurate. This solution achieved an average of 83\% of the gains obtained with instrumentation-based PGO and 93\% on C++ benchmarks only. The profiling overhead is only 1.06\% on average whereas instrumentation incurs a 16\% overhead on average.

\vspace{5mm}


\end{abstract}

\section{Introduction}

Profilers help developers and compilers find the main areas for optimization. Profiling and optimizing by hand is a time-consuming process, but this process can be automated. Modern compilers include an optimization technique called Profile-Guided Optimization (PGO). Feedback-Directed Optimization (FDO) is also used as a synonym of PGO.

PGO uses information collected during the execution of a program to optimize it. Generally, edge execution frequencies between basic blocks are collected. Several optimization techniques can take advantage from the collected profile.  For instance, the data can be used to drive inlining decisions and block ordering within a function to achieve minimal cacheline usage. Branches can be reordered based on their frequency to avoid branch misprediction. Loops working on arrays causing Data Cache misses can be improved to make better use of the cache. As the dynamic profile, unlike the static profile, captures execution frequency, this can result in impressive speedups for non IO-intensive applications.

Compilers currently support instrumentation-based PGO. In this variant, the compiler must first generate a special version of the application in which instructions are inserted at specific locations to generate the profile. During the execution, counters are incremented by these instructions and finally, the profile is generated into a file. After that, the program is compiled again, this time with PGO flags, to use the profile and optimize the binary for the final version.

This approach has several drawbacks:

\begin{itemize}
    \item The instructions inserted into the program slow it down. An instrumented binary can be much slower than its optimized counterpart. This has been reported to incurs between 9\% and 105\% overhead\cite{Ball94,Ball96}. In practice, it has been observed to be as much as an order of magnitude slower for some applications.
    \item The profile data must be collected on a specially compiled executable. This dual-compilation model is not convenient. Indeed, for applications with long build time, doubling this time may degrade productivity.
    \item Only a small set of information can be collected. For example, it is not possible with this approach to collect information about memory or branch-prediction issues.
    \item There is a tight coupling between the two builds. It is generally necessary to use the same optimization options both in the first and second compilation. Without that, the control-flow graph (CFG) of both compilation may not match and the profiling data may not be used with enough accuracy. Making large changes in the source code also invalidates the previous profile data.
    \item The instrumentation instructions can alter the quality of the generated profile. As new instructions are inserted into the program, they may change the results of the profiling, which may, in turn change the optimization decisions.
\end{itemize}

For these reasons, traditional PGO has not been widely adopted. Even most of the CPU-intensive projects are not using this technique. To avoid these drawbacks, we propose in this paper a solution based on sampling Hardware Performance Events generated by the Performance Monitoring Unit (PMU) instead of instrumenting the application. Source position contained in the debug symbols is used to recreate an accurate profile.

Below, we list the primary contributions of this work:
\begin{enumerate}
    \item We study Hardware Performance Events and their use for PGO.
    \item We build a complete toolchain, based on GCC, able to perform sampling-based PGO.
    \item Finally, we evaluate the performance of our implementation. We present results obtained with our implementation in the GCC compiler with SPEC 2006 benchmarks. We show that our toolchain can achieve 93\% of the gains obtained using instrumentation-based PGO and incurs only a 1.06\% overhead, where instrumentation adds 16\% overhead on average.
\end{enumerate}

The rest of this paper is organized as follows: Section~\ref{sec:pgo-counters} describes how to combine PGO and sampling Hardware Performance Events. Section~\ref{sec:implementation} then presents the toolchain that has been developed. Section~\ref{sec:results} describes the results obtained with sampling-based PGO. Section~\ref{sec:related} lists related work in the area. Finally, Section~\ref{sec:conclusion} presents our conclusions and future work for this project.

\section{PGO and Performance Counters}
\label{sec:pgo-counters}

This section describes how sampling Hardware Performance Counters can be used to perform PGO.

\subsection{Hardware Performance Events}

Every modern microprocessor includes a set of counters that are updated when a particular hardware event arises. This set of counters is managed by the Performance Monitoring Unit (PMU) of the CPU. These events can then be accessed by an other application.

The most common way of using these counters is by sampling. The PMU has to be configured to generate an interrupt when a particular counter goes over a specific limit (it overflows). At this point, the monitoring software can record the state of the system and especially the current executed instruction, indicated by the Program Counter (PC). This directly generates a complete instruction profile for the binary instructions.

A basic block profile can be naturally estimated from sampling. Each time the counter overflows, the instruction identified by the PC is saved. At the end of the execution, the number of samples for each instruction of a basic block are summed. The basic block sums must be normalized to avoid giving higher weight to larger basic blocks.

There is a lot of different events, from clock cycles, to L2 cache misses or branch mispredictions. The available events depend on the microarchitecture. Some processors provides very large list of events (more than 1,500 for the PowerPC Power7 family) and some much fewer (69 for the ARM V7 Cortex-A9 processor).

The solution presented in this paper has been specially tuned for Intel\textsuperscript{\textregistered} Core\textsuperscript{\texttrademark} i7 events\cite{Levinthal2009}.

\subsection{Sampling-Based PGO}

Combining PGO and Hardware Performance Counters results in the sampling-based PGO technique. In this model, the program that is profiled is directly the production binary, there is no need for a special executable. However, the profile is this time generated with a specific program that can sample the values of the Hardware Performance Counters, namely a profiler. Once a profile is generated, the compiler can use it for the next compilations.

The main advantage of this approach is the much smaller overhead of sampling compared to instrumentation. The program is only interrupted when a counter overflows, not every time a function is executed for instance. The cost of sampling depends on the sampling period and on the event that is sampled.

Moreover, the profiler can be patched on an already running program. It means that the production executable can be profiled for some hours without interrupting it. Profiling on the production binary, with the production input data, generally results in more accurate profiles. Moreover, it is not necessary to find training data for the instrumentation executable, a task which can be hard depending on the profiled execution.

Since source position is used to match the program and the profile, the coupling between the profile and the binary is much smaller. Changing the compilation options would not invalidate the profile. Moreover, older profiles can still be used in new versions of the application. It is not necessary to generate profiles for each version of the program, except in the case of major changes in the application.

The data generated by the profiler must be transformed before being used for PGO. The samples contain only the address of the instructions and the sample count. Information such as the source location of the instruction is necessary to generate a real instruction profile useful for the compiler from this raw data. This is explained in details in Section~\ref{sec:gathering-profile}

On the other hand, there are also some drawbacks to this method:
\begin{itemize}
    \item As the supported events are depending on the microarchitecture, sampling very specific events may not be portable
    \item As not all events are recorded, this method is not as accurate as instrumentation-based profile. In practice, it showed to be accurate enough for bringing performance speedups.
    \item The sampling period must be chosen carefully. Indeed, a longer sampling period means a larger overhead, but a more accurate profile. It is important to find the correct balance between the two factors. After a certain point, it is not really interesting to sample more events.
    \item It may occur that the sampling period is synchronized with some part of the code which can lead to the point where only one instruction of a loop is reported \cite{Chen2010}. This problem can be solved by using randomization on the period.
\end{itemize}

Even with these drawbacks, sampling-based PGO is a better fit in the development process than the traditional approach.

\section{Implementation}
\label{sec:implementation}

Our implementation relies on the following components:

\begin{enumerate}
    \item perf\cite{DeMelo2010}: The Hardware Performance Counters profiler of the Linux kernel.
    \item Gooda\footnote{\url{https://code.google.com/p/gooda/}}\cite{Calafiura2012}: An analyzer for perf data files. It is the result of a collaboration between the Lawrence Berkeley National Laboratory (LBNL) and Google. %The main author of this tool being David Levinthal.
    \item AutoFDO\footnote{\url{http://gcc.gnu.org/wiki/AutoFDO}} (AFDO): A patch from Google bringing sampling-based PGO to GCC. See Section~\ref{sec:afdo}
\end{enumerate}

In this implementation, PGO is made in several steps (the first three steps being automated in one for the user by the profile generator):

\begin{enumerate}
    \item The application is profiled with perf. A certain set of events is collected.
    \item Gooda generates intermediate spreadsheets by analyzing the perf data files.
    \item The spreadsheets are converted into an AFDO profile by our implementation.
    \item GCC reads the profile with the AFDO patch and optimizes the code.
\end{enumerate}

\begin{figure}
    \centering
    \includegraphics[scale=0.9]{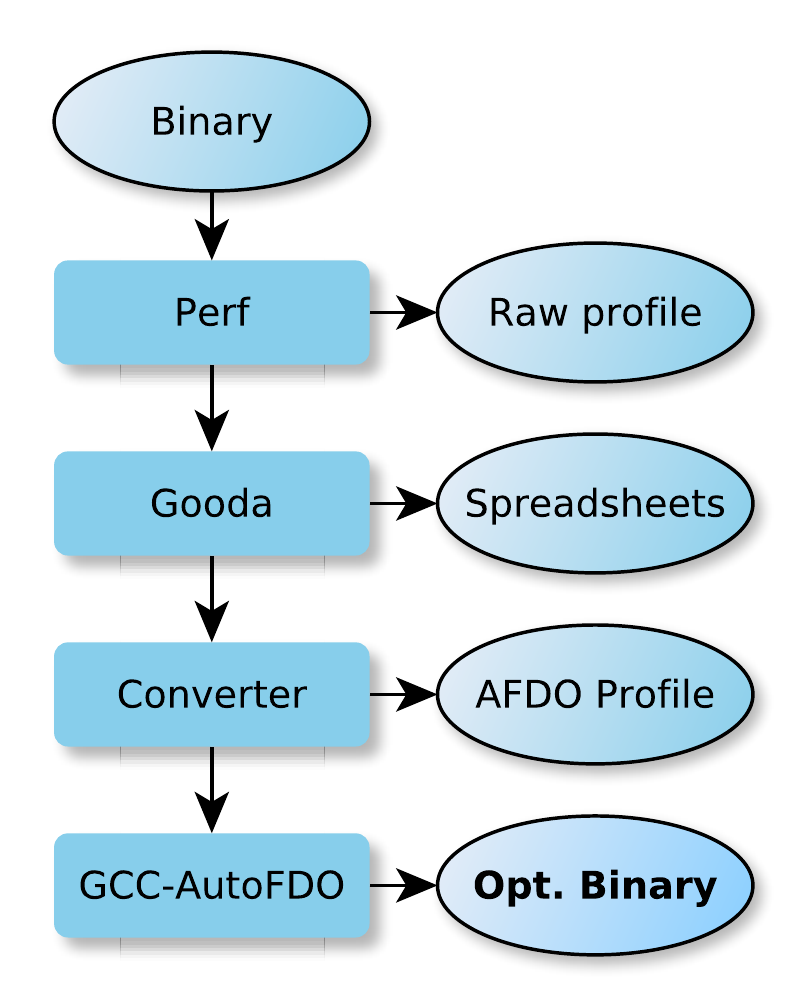}
    \caption{Profile process}
    \label{fig:machinery}
\end{figure}

Figure~\ref{fig:machinery} shows an overview of these steps and the intermediate files that are used in the process.

The profile used by this toolchain is an instruction profile, there is a counter value for every instruction of the application. This profile does not comprehend basic blocks. The basic block profile is computed from the instruction profile inside GCC by AFDO. This profile also includes the entry count and execution count for each function.

Two modes have been implemented (Cycles Counting and LBR), see Section~\ref{seg:cycles-mode} and Section~\ref{seg:lbr-mode}. In both modes, an instruction profile is generated with a counter for each instruction.

Gooda being able to merge several spreadsheets, it is possible to collect profiles on several machines of a cluster for instance and then combine all of them to have an accurate global profile. This can also be used when the same executable is run with different data sets to merge the resulting profiles.

\subsection{AutoFDO}
\label{sec:afdo}

AutoFDO (AFDO)\cite{Chen2013} is a patch for GCC developed by Google
%TODO and especially, Dehao Chen
. AFDO is the successor of SampleFDO \cite{Chen2010}. It has been rewritten from scratch and has several major differences. SampleFDO was reconstructing the CFG for each function using a Minimum Control Flow (MCF) algorithm. It uses a complex two-phase annotation pass to handle inlined functions. These two techniques were very expensive and complicated, therefore they have been abandoned in AFDO.

AFDO uses source profile information. The profile maps each instruction to an inline stack, itself mapped to runtime information. An inline stack is a stack of functions that have been inlined at a specific call site. The execution count and the number of instructions mapped for each source instruction are collected at runtime. Debug information are used to reconstruct the profile from the runtime profile. While AFDO has been developed for \texttt{perf}, it is independent from it. The profile could be generated from another hardware events profiler.

AFDO is activated using a command-line switch. A special pass is made to read the profile and load it into custom data structures. The first special use of the profile data is made during the \texttt{early\_inline} pass. To make the profile annotation easier, AFDO ensures that each hot call site that was inlined in the profiled binary is also inlined during this pass. For this, a threshold-based top-down decision is used, during a bottom-up traversal of the call graph.

Once early inlining decisions have been made, the execution profile is annotated using the AFDO profile. The basic block counts are directly taken from the profile, whereas the edge counts are propagated from the execution counts. The strength of AFDO is that it already profits from all the GCC optimization that are using the profile. When PGO is used, the static profile used in optimization passes is replaced with a profile generated from the AFDO profile. All backend optimization passes use profile information just as normally.

However, some special tuning still needs to be done. For instance, during Interprocedural Register Allocation (IRA), the function is considered to not be called if its entry basic block's count is zero. However, in sampling-based PGO, this is not necessary the case and special care is taken to ensure that function is not considered dead. The same problem arises in the hot call site heuristic where the entry count of the callee function is checked for zero. In this case, the heuristic is disabled if it is zero.

AFDO is especially targeting C/C++ applications and therefore is specially tuning callgraph optimizations.

\subsection{Cycles Counting}
\label{seg:cycles-mode}

In this first mode, the counter that is used is the number of Unhalted Core Cycles for each instruction. It is a common measure of performance, as it computes only the time when the processor is doing something, not when it is waiting for another operation, I/O for instance.

This mode is based on common Cycle Accounting Analysis for Intel processors\cite{Levinthal2008}.

In this mode, the instruction profile is naturally generated as events are directly mapped to an instruction.

\subsection{LBR Mode}
\label{seg:lbr-mode}

To have a better accuracy, the second mode uses the Last Branch Record (LBR) feature of the PMU.

The LBR is a trace branch buffer. It captures the source and target addresses of each retired taken branch. It can track 16 pairs of addresses. It provides a call tree context for every event for which LBR is enabled.

In this implementation, we do not directly manipulate the LBR data. We take advantage of the fact that Gooda already merges together the LBR samples to compute the number of basic block executions.

The counter used in LBR mode is the number of Branch Instruction Retired\footnote{an instruction is retired when it has been executed and its result is used}. It is an interesting event because it makes it easy to compute the number of execution of each basic-block as it references each branch. By using the 16 addresses of the LBR history, the basic block paths can be computed with high accuracy.

The instruction profile is generated from the basic block profile, every instruction of a basic block having the same counter value.

\subsection{Gathering the profile}
\label{sec:gathering-profile}

The profile generated by perf and preprocessed by Gooda is a binary instruction profile. It means that each instruction is identified by its address. This address is not useful inside GCC, because during the optimization process, instructions are not assigned an address. To identify an instruction inside GCC, it is necessary to have its source position.

Each binary instruction address must be mapped to its source location. For that, the debug symbols are extracted from the ELF executable and then are used to reconstruct the source profile..

Each instruction is identified by four different values: its filename, the containing function name, its line number and its discriminator.

The DWARF discriminator allows to discriminate different statements on the same line. This is very common in modern programming languages. Discriminators are gathered on the executable using \texttt{addr2line} for each instruction.

Another important point to distinguish instructions is the handling of inlined functions. When a function gets inlined, there is a copy of the instructions of the called function at several places in the binary. All those different copies have the same source location debug symbols, so they are not enough to distinguish them. Fortunately, DWARF symbols include the inline stack for each inlined function. By storing the inline stack of each instruction, the profile is very accurate and provides a correct mapping between binary instructions and source instructions. These inline stacks are then processed directly by AFDO. This processing proved very important on large C++ projects.

The last point of importance concerning the creation of the profile is the function name. The name of a function is generally not enough to identify it uniquely. When using GCC, a function has two more names: the mangled (or assembler) name identifies uniquely a function (take into account all the parameters) and the Binary File Descriptor (BFD) name, used by GCC in the debug symbols to identify functions in inline stack. To identify each source function, its assembler name is taken from the table of symbols of the ELF file.

\subsection{Shortcomings}

The present implementation has some limitations. First of all, the debug symbols are absolutely essential in order to reconstruct the correct instruction profile. Indeed, without debug symbols, it would not be possible to match assembly instructions to their original source locations. Our implementation works only for programs that have been compiled in debug mode, but optimizations can be enabled. Debug executables are not slower than stripped executables, but they can be much bigger. This can be overcome by keeping a stripped copy of the optimized and use it on production after PGO. However, the production binary could not be used for profiling.

Moreover, if the compiler does not generate precise debug symbols, the profile will not be accurate. This is especially a problem as some optimizations are not preserving correctly the debug symbols. Another problem comes in the form of over or under sampling. For instance, Loop Unrolling may cause the same statements to be duplicated in different basic blocks. Normalization may then lead to a profile too low for the generated basic blocks.

It is also highly depending on \texttt{addr2line} and \texttt{objdump} to gather information about the binary file. To have all the features, it is necessary to  possess a very recent version of binutils, at least 2.23.1. Moreover, if there is a bug in either of these tools, the profile generated could be inaccurate.

Another shortcoming comes from the fact that the DWARF debugging data format does not support discriminators in inline stacks. It means that the profile is not completely precise and it can lead the compiler to the wrong decision, even if this has not been found to be an issue in practice.

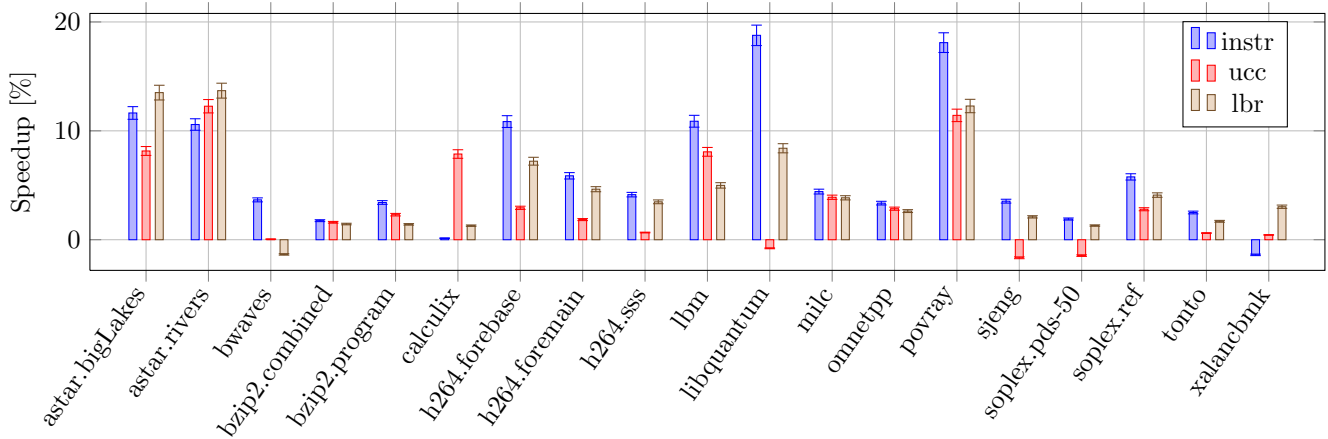
\begin{figure*}[ht]
    \begin{tikzpicture}
        \begin{axis}[
                width=18cm,
                height=5cm,
                ybar,
                bar width=3pt,
                symbolic x coords={astar.bigLakes, astar.rivers, bwaves, bzip2.combined, bzip2.program, calculix, h264.forebase,h264.foremain,h264.sss,lbm,libquantum,milc,omnetpp,povray,sjeng,soplex.pds-50,soplex.ref,tonto,xalancbmk},
                xtick=data,
                legend pos=north east,
                x tick label style={rotate=55,anchor=east},
                ymajorgrids=true,
                xmajorgrids=true,
                ylabel=Speedup,
                y unit=\%,
                enlargelimits=0.05
            ]

            \addplot+[error bars/.cd,y dir=both,y fixed relative=0.05] coordinates {(astar.bigLakes,11.64) (astar.rivers,10.58) (bwaves,3.66) (bzip2.combined,1.75) (bzip2.program,3.42) (calculix,0.14) (h264.forebase,10.85) (h264.foremain,5.87) (h264.sss,4.14) (lbm,10.88) (libquantum,18.77)(milc,4.42)
            (omnetpp,3.36)(povray,18.10)(sjeng,3.54)(soplex.pds-50,1.90)(soplex.ref,5.77)(tonto,2.50)(xalancbmk,-1.39)};
            \addplot+[error bars/.cd,y dir=both,y fixed relative=0.05] coordinates {(astar.bigLakes,8.15) (astar.rivers,12.26) (bwaves,0.05) (bzip2.combined,1.60) (bzip2.program,2.30) (calculix,7.87) (h264.forebase,2.93) (h264.foremain,1.85) (h264.sss,0.66) (lbm,8.07) (libquantum,-0.79)(milc,3.90)
            (omnetpp,2.85)(povray,11.42)(sjeng,-1.66)(soplex.pds-50,-1.46)(soplex.ref,2.80)(tonto,0.61)(xalancbmk,0.44)};
            \addplot+[error bars/.cd,y dir=both,y fixed relative=0.05] coordinates {(astar.bigLakes,13.51) (astar.rivers,13.69) (bwaves,-1.35) (bzip2.combined,1.44) (bzip2.program,1.41) (calculix,1.29) (h264.forebase,7.21) (h264.foremain,4.64) (h264.sss,3.48) (lbm,4.99) (libquantum,8.40)(milc,3.85)
            (omnetpp,2.63)(povray,12.28)(sjeng,2.10)(soplex.pds-50,1.30)(soplex.ref,4.10)(tonto,1.69)(xalancbmk,3.03)};

            \legend{instr,ucc,lbr}
        \end{axis}
    \end{tikzpicture}
    \caption{Speedups for SPEC CPU 2006 benchmarks. The application is trained with training data set. Our implementation achieves 75\% of instrumented PGO.}
        \label{fig:all-results-training}
    \end{figure*}

\section{Experimental Results}
\label{sec:results}

The implementation has been evaluated in terms of speedups compared to the optimized version and to the instrumentation-based PGO version. All binaries were produced using a trunk version of the Google GCC 4.7 branch. The target was an x86\_64 architecture. The processor used for the tests was an Intel\textsuperscript{\textregistered} Xeon\textsuperscript{\texttrademark} E5-2650, 2 GHz.

Four versions are compared:
\begin{itemize}
    \item \texttt{base}: The optimized executable, compiled with \texttt{-O2 -march=native}. The same flags are also used for the other executables with additional flags.
    \item \texttt{instr}: The executable trained with instrumentation-based PGO.
    \item \texttt{ucc}: The program trained with our implementation in Cycle Accounting mode. \texttt{UNHALTED\_CORE\_CYCLES} is sampled with a period of 2'000'000 events.
    \item \texttt{lbr}: The program trained with our implementation in LBR mode. \texttt{BRANCH\_INST\_RETIRED} is sampled with a period of 400'000 events.
\end{itemize}

All the results were collected on the SPEC CPU 2006 V1.2 benchmarks\footnote{\url{http://www.spec.org/cpu2006/}}. Each version of the binary is run three times and SPEC choose the best result out of the three runs. The variation between the runs is very low.

For run time reasons, a subset of the benchmarks has been selected. This subset has been chosen to be representative of different programming languages, both floating points and integers programs and to represent different fields of science.

For these tests, a modified Gooda version was used to work on small programs. Gooda is especially made for big programs and so the profile is not generated for functions below some thresholds. It has been necessary to change the thresholds in order to include as much functions as possible in the profile.

Figure~\ref{fig:all-results-training} shows the performance of the different PGO versions. Each version is compared to the performance of the \texttt{base} version.

The results are very interesting. Indeed, LBR achieves about 75\% of the instrumentation-based PGO gains (arithmetic average of percentages). Cycle Accounting is less effective, but still achieves 53\% of the gains. This was expected as LBR should be more accurate than Unhalted Core Cycles.

In some cases, sampling-based PGO even outperforms the traditional approach. For instance, on astar benchmarks, LBR achieves 116\% to 129\% of the instrumentation gains. This is even more true for calculix and xalancbmk where instrumentation-based PGO performs very poorly and sampling achieves good results. This difference is not so surprising, as several optimizations are driven by threshold based heuristics, so small differences in the profile can drastically change decisions and lead to better (or worse) performances.

On the contrary, there are also benchmarks where our implementation performs poorly compared to traditional PGO. For instance, bwaves proved a very bad case for our implementation. It comes from the fact that AutoFDO is not optimized for Fortran code, since it has been developed with a C/C++ centric approach. Indeed, special care has been taken to tune inlining and call optimization passes, while tuning for loop-intensive code has not been performed.

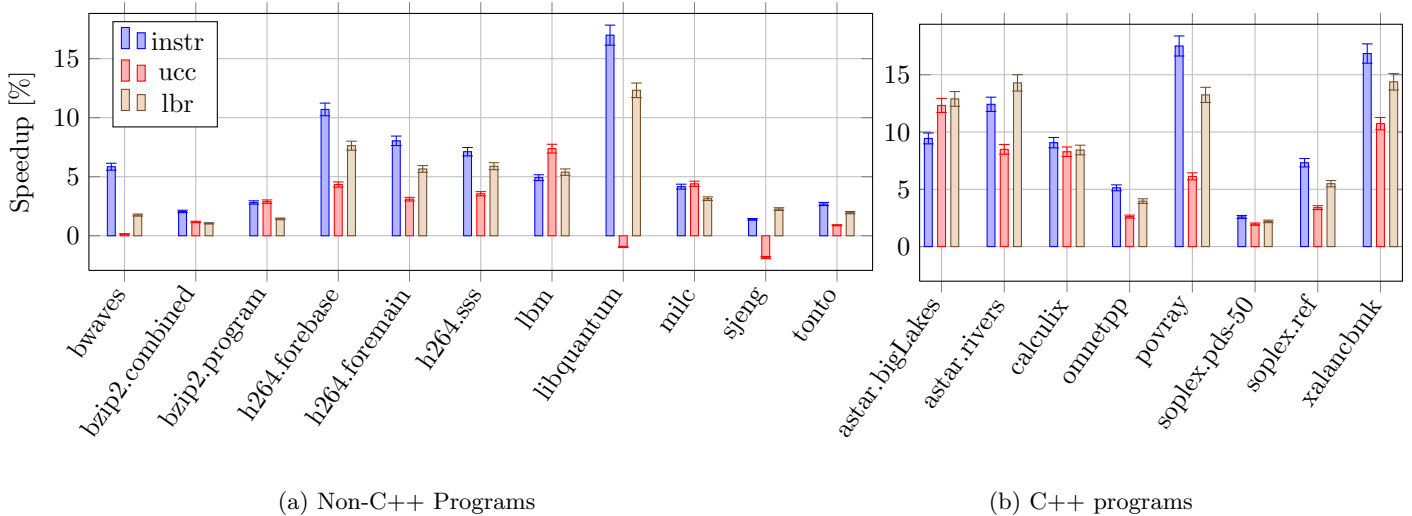
\begin{figure*}
    \begin{subfigure}[b]{0.6\textwidth}
        \begin{tikzpicture}
            \begin{axis}[
                    width=12cm,
                    height=5cm,
                    ybar,
                    bar width=3pt,
                    symbolic x coords={bwaves, bzip2.combined, bzip2.program, h264.forebase,h264.foremain,h264.sss,lbm,libquantum,milc,sjeng,tonto},
                    xtick=data,
                    legend pos=north west,
                    x tick label style={rotate=55,anchor=east},
                    ymajorgrids=true,
                    xmajorgrids=true,
                    ylabel=Speedup,
                    y unit=\%,
                    enlargelimits=0.05
                ]
                \addplot+[error bars/.cd,y dir=both,y fixed relative=0.05] coordinates {(bwaves,5.85) (bzip2.combined,2.06) (bzip2.program,2.82) (h264.forebase,10.70) (h264.foremain,8.04) (h264.sss,7.12) (lbm,4.92) (libquantum,16.99)(milc,4.16) (sjeng,1.39)(tonto,2.69)};
                \addplot+[error bars/.cd,y dir=both,y fixed relative=0.05] coordinates {(bwaves,0.15) (bzip2.combined,1.17) (bzip2.program,2.89) (h264.forebase,4.34) (h264.foremain,3.10) (h264.sss,3.56) (lbm,7.38) (libquantum,-0.95)(milc,4.40) (sjeng,-1.85)(tonto,0.89)};
                \addplot+[error bars/.cd,y dir=both,y fixed relative=0.05] coordinates {(bwaves,1.75) (bzip2.combined,1.05) (bzip2.program,1.43) (h264.forebase,7.63) (h264.foremain,5.66) (h264.sss,5.89) (lbm,5.39) (libquantum,12.32)(milc,3.14) (sjeng,2.26)(tonto,1.96)};
                \legend{instr,ucc,lbr}
            \end{axis}
        \end{tikzpicture}
        \caption{Non-C++ Programs}
        \label{fig:non-cpp-results-reference}
    \end{subfigure}
    \begin{subfigure}[b]{0.4\textwidth}
        \begin{tikzpicture}
            \begin{axis}[
                    width=8cm,
                    height=5cm,
                    ybar,
                    bar width=3pt,
                    ymin=-2,
                    symbolic x coords={astar.bigLakes, astar.rivers, calculix, omnetpp,povray,soplex.pds-50,soplex.ref,xalancbmk},
                    xtick=data,
                    x tick label style={rotate=55,anchor=east},
                    ymajorgrids=true,
                    xmajorgrids=true,
                    enlargelimits=0.05
                ]
                \addplot+[error bars/.cd,y dir=both,y fixed relative=0.05] coordinates {(astar.bigLakes,9.44) (astar.rivers,12.42) (calculix,9.07) (omnetpp,5.13)(povray,17.51)(soplex.pds-50,2.59)(soplex.ref,7.32)(xalancbmk,16.85)};
                \addplot+[error bars/.cd,y dir=both,y fixed relative=0.05] coordinates {(astar.bigLakes,12.32) (astar.rivers,8.49) (calculix,8.28) (omnetpp,2.62)(povray,6.13)(soplex.pds-50,1.96)(soplex.ref,3.40)(xalancbmk,10.73)};
                \addplot+[error bars/.cd,y dir=both,y fixed relative=0.05] coordinates {(astar.bigLakes,12.89) (astar.rivers,14.29) (calculix,8.43) (omnetpp,3.97)(povray,13.24)(soplex.pds-50,2.20)(soplex.ref,5.49)(xalancbmk,14.38)};
            \end{axis}
        \end{tikzpicture}
        \caption{C++ programs}
        \label{fig:cpp-results-reference}
    \end{subfigure}
    \caption{Speedups for SPEC CPU 2006 benchmarks. The application is trained with reference data set. Our implementation achieves 84\% of instrumented PGO on overall benchmarks and 93\% on C++ programs only.}
    \label{fig:all-results-reference}
\end{figure*}

The first results were obtained by training the executable on the training data set. To see the impact of the input data set, the executables were then trained again on the reference data set. It means that the same input data set is used for training and for benchmarking the executable. Figure~\ref{fig:all-results-reference} shows the speedups obtained when training on the reference data set.

This time, LBR is able to achieve 84\% of the instrumentation gains. An interesting point about these results is that where \texttt{instr} and \texttt{ucc} improve their scores by about 22\%, \texttt{lbr} improves by 37\%. LBR-sampling seems to be even more interesting when the input data is closer to the real data set. Most of the benchmarks improved only a bit with the reference data set, but xalancbmk improved by an order of magnitude. It seems that in its case, the training data set is not close enough to the reference data set for PGO. calculix seems to be in the same case, although the difference is not so spectacular.

As AFDO has been especially tuned for C++, Figure~\ref{fig:cpp-results-reference} presents the results for C++ benchmarks only.

\begin{figure}
\end{figure}

On C++ benchmarks, both sampling versions are performing very well. Cycle Accounting achieves 67\% of the instrumentation gains and LBR reaches 93\%. These results are very promising. For each benchmark, \texttt{lbr} is very close (and sometimes better) to \texttt{instr}.

This section presented some results that can still be improved, especially for some non-C++ benchmarks. Once AFDO has improved support for other languages, like Fortran, it may be interesting to run these tests again to see how much of the instrumentation gains sampling-based PGO can reach. It may also be interesting to investigate the benchmarks where \texttt{lbr} proved better than \texttt{instr} and see if the result can be obtained on other benchmarks as well.

\subsection{Sampling Overhead}

The overhead of our implementation has also been evaluated and compared to the overhead of instrumentation-based PGO.

Four versions are compared:
\begin{itemize}
    \item \texttt{base}: The optimized executable, compiled with \texttt{-O2 -march=native}.
    \item \texttt{instr}: The GCOV instrumented executable.
    \item \texttt{ucc:} The program run under the monitoring of the profiler with sampling on \texttt{UNHALTED\_CORE\_CYCLES} with a period of 2'000'000 events.
    \item \texttt{lbr}: The program run under the monitoring of the profiler \texttt{BRANCH\_INST\_RETIRED} with a period of 400'000 events.
\end{itemize}

The sampling periods used for this test have been chosen by empiric evaluation and chosing the best one, based on the speedup result and the overhead.

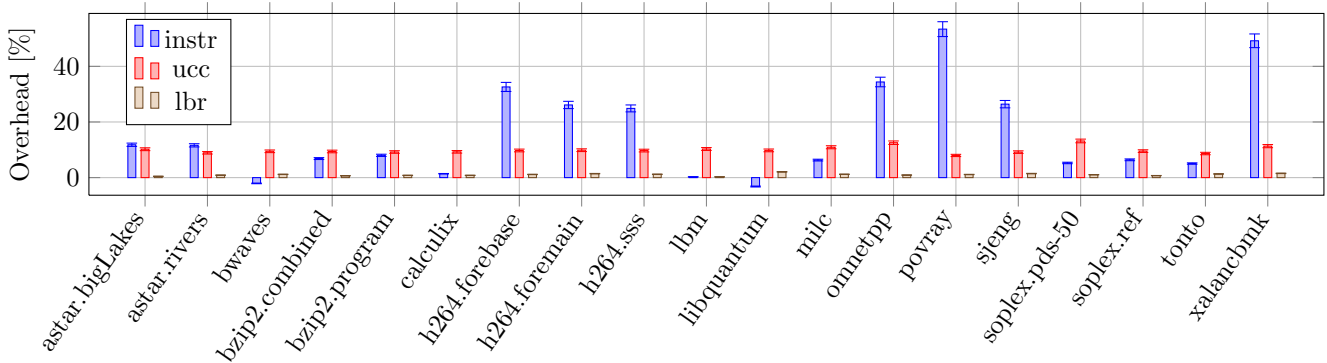
\begin{figure*}
    \begin{tikzpicture}
        \begin{axis}[
                width=18cm,
                height=4cm,
                ybar,
                bar width=3pt,
                symbolic x coords={astar.bigLakes, astar.rivers, bwaves, bzip2.combined, bzip2.program, calculix, h264.forebase,h264.foremain,h264.sss,lbm,libquantum,milc,omnetpp,povray,sjeng,soplex.pds-50,soplex.ref,tonto,xalancbmk},
                xtick=data,
                legend pos=north west,
                x tick label style={rotate=55,anchor=east},
                ymajorgrids=true,
                xmajorgrids=true,
                ylabel=Overhead,
                y unit=\%,
                enlargelimits=0.05
            ]

            \addplot+[error bars/.cd,y dir=both,y fixed relative=0.05] coordinates {(astar.bigLakes,11.81) (astar.rivers,11.61) (bwaves,-2.09) (bzip2.combined,6.85) (bzip2.program,8.01) (calculix,1.41) (h264.forebase,32.60) (h264.foremain,26.12) (h264.sss,24.88) (lbm,0.26) (libquantum,-3.20)(milc,6.30)
            (omnetpp,34.38)(povray,53.38)(sjeng,26.41)(soplex.pds-50,5.27)(soplex.ref,6.40)(tonto,5.06)(xalancbmk,49.19)};
            \addplot+[error bars/.cd,y dir=both,y fixed relative=0.05] coordinates {(astar.bigLakes,10.21) (astar.rivers,8.92) (bwaves,9.46) (bzip2.combined,9.40) (bzip2.program,9.21) (calculix,9.27) (h264.forebase,9.76) (h264.foremain,9.86) (h264.sss,9.72) (lbm,10.29) (libquantum,9.79)(milc,10.92)
            (omnetpp,12.55)(povray,7.97)(sjeng,9.17)(soplex.pds-50,13.18)(soplex.ref,9.53)(tonto,8.66)(xalancbmk,11.32)};
            \addplot+[error bars/.cd,y dir=both,y fixed relative=0.05] coordinates {(astar.bigLakes,0.46) (astar.rivers,0.87) (bwaves,1.17) (bzip2.combined,0.68) (bzip2.program,0.81) (calculix,0.81) (h264.forebase,1.14) (h264.foremain,1.43) (h264.sss,1.19) (lbm,0.26) (libquantum,2.08)(milc,1.19)
            (omnetpp,0.91)(povray,1.12)(sjeng,1.46)(soplex.pds-50,1.05)(soplex.ref,0.73)(tonto,1.32)(xalancbmk,1.53)};

            \legend{instr,ucc,lbr}
        \end{axis}
    \end{tikzpicture}
    \caption{Overheads for SPEC CPU 2006 benchmarks. Our implementation has 15 times less overhead than instrumentation-based PGO.}
    \label{fig:overhead-all-results}
\end{figure*}

Figure~\ref{fig:overhead-all-results} shows the overhead of the different versions, compared to the \texttt{base} version.

The overhead of instrumentation is high, but not as high as expected, only 16\% on average. The highest overhead of instrumentation was 53\% on the povray benchmark.

As expected, the overhead of sampling is lower than the overhead of instrumentation. LBR has an average of 1.06\% of overhead, which is 15 times lower than instrumentation. In several benchmarks the overhead is less than 1\%.

Unfortunately, the overhead of Cycle Counting is much higher than it should be, indeed, it is as high as 10\% in average. The problem is that, in this mode, Gooda profiles a large number of events, not only Unhalted Core Cycles. This adds a large overhead in terms of performance. At this time, it is not possible to configure Gooda to only use Unhalted Core Cycles. On the other hand, as LBR event proved more accurate and has a small overhead, the value of this mode is reduced.

Another point of interest is that the variability of the overhead is much higher for instrumentation than for sampling. Indeed, the overhead of instrumentation varies from 0\% to 53\%, whereas the overhead of LBR-sampling varies only from 0.3\% to 2\%. The instrumentation overhead highly depends on the code that is being benchmarked. On the other hand, sampling overhead depends mostly on the period of sampling and the type of the sampled events.

Some instrumented executables proved faster than their optimized counterpart. It may happen in some cases. After the instrumentation instructions are inserted in the program, the program is optimized further. These new instructions may change the decisions that are made during the optimization process. Such changes in decision may lead to faster code. This situation cannot happen with sampling.

\subsection{Tools Overhead}

In the previous section, only the overhead of sampling (with \texttt{perf}) has been measured. The time necessary to generate the profile from perf also needs to be taken into accounts.

There are two different tools adding overhead to the overall process. The first one, also the slowest one, being Gooda. Until now, Gooda has not been tuned for performances and it may be quite slow for handling very large profiles. The conversion from Gooda spreadsheets to an AFDO profile is not so critical, since Gooda already filters several functions. Moreover, it has already been tuned for performances so as to add the smallest possible overhead to the process.

Both overhead have been tested on several profiles gathered using perf in cycle accouting mode (\texttt{UNHALTED\_CORE\_CYCLES} with a period of 2'000'000 events). The profiles have been gathered on GCC compiling two different programs, a toy compiler and the converter itself. The perf profile are varying from 167MiB (\texttt{gcc-google-eddic}) and 194KiB (\texttt{eddic-list}). Each test has been run five times and the best result has been taken. The variations between different runs was very low.

\begin{figure}
    \begin{subfigure}[b]{0.23\textwidth}
        \begin{tikzpicture}
            \begin{axis}[
                    width=4.5cm,
                    height=4cm,
                    ybar,
                    bar width = 4pt,
                    symbolic x coords={gcc-converter,gcc-eddic,eddic-assembly,eddic-list,converter-ucc,converter-lbr},
                    xtick=data,
                    x tick label style={rotate=55,anchor=east},
                    ymajorgrids=true,
                    xmajorgrids=true,
                    ylabel=Overhead,
                    y unit=\%,
                    enlargelimits=0.05,
                    ymin=0
                ]
                \addplot coordinates {(gcc-converter,3.9) (gcc-eddic,3.601)(eddic-assembly,6.003)(eddic-list,4.415)(converter-ucc,0.742)(converter-lbr,0.620)};
            \end{axis}
        \end{tikzpicture}
        \caption{Converter Overhead}
        \label{fig:converter-tools-overhead}
    \end{subfigure}
    \begin{subfigure}[b]{0.23\textwidth}
        \begin{tikzpicture}
            \begin{axis}[
                    width=4.5cm,
                    height=4cm,
                    ybar,
                    bar width = 4pt,
                    symbolic x coords={gcc-converter,gcc-eddic,eddic-assembly,eddic-list,converter-ucc,converter-lbr},
                    xtick=data,
                    x tick label style={rotate=55,anchor=east},
                    ymajorgrids=true,
                    xmajorgrids=true,
                    enlargelimits=0.05,
                    ymin=0
                ]
                \addplot coordinates {(gcc-converter,87.651)(gcc-eddic,121.419)(eddic-assembly,7.192)(eddic-list,5.702)(converter-ucc,3.748)(converter-lbr,2.797)};
            \end{axis}
        \end{tikzpicture}
        \caption{Gooda Overhead}
        \label{fig:gooda-tools-overhead}
    \end{subfigure}
    \caption{Overhead of the profile generation.}
\end{figure}
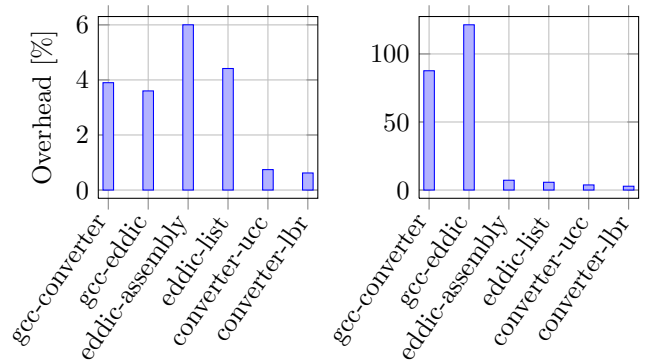

Figure~\ref{fig:converter-tools-overhead} presents the overhead of the converter. As shown in this figure, the overhead of the converter is not negligible. It takes a maximum of six seconds for the test cases. It has to be put in regard of the running time of the profiling. For instance, the test case \texttt{gcc-eddic} runs during 40 minutes. This makes an overhead of 0.25\%, which is acceptable. An important point to consider is that it does not scale directly with the size of the profile but with the number of functions reported by Gooda, which should grow up to a maximum related to the size of the profiled executable. In the converter, about 65\% of the time is spent in calling \texttt{addr2line}. This could be improved by integrating address to line conversions directly in the converter.

Figure~\ref{fig:gooda-tools-overhead} shows the overhead of Gooda, converting the perf profile to a set of spreadsheets. It is very clear that the overhead of Gooda is much higher than the overhead of the converter and is considerable for at least two test cases (both gcc test cases). On the slowest test case (\texttt{gcc-eddic}), the overhead is as high as five percent, which makes the tool chain much less interesting. It also takes several seconds for the other samples even if they are much faster to run under profiling. The overhead is generally getting better with the running time of the program. In its current state, the current toolchain is more adapted to long-running programs.

The long running time of Gooda is something that should really be improved in the future.

\section{Related Work}
\label{sec:related}

In 2008, Roy Levin, Ilan Newman and Gadi Haber \cite{Levin2008} proposed a solution to generate edge profiles from instruction profiles of the instruction retired hardware event for the IBM FDPR-Pro, post-link time optimizer. This solution works on the binary level. The profile is applied to the corresponding basic blocks after link-time.  The construction of the edge profile from the sample profile is known as a Minimum Cost Circulation problem. They showed that this can be solved in acceptable time for the SPEC benchmarks, but this remains a heavy algorithm.

Soon after Levin et al., Vinodha Ramasamy, Robert Hundt, Dehao Chen and Wenguang Chen \cite{Ramasamy2008} presented another solution of using instruction retired hardware events to construct an edge profile. This solution was implemented and tested in the Open64 compiler. Unlike the previous work, the profile is reconstructed from the binary using source position information. This has the advantage that the binary can be built using any compiler and then used by Open64 to perform PGO.  They were able to reach an average of 80\% of the gains that can be obtained with instrumentation-based PGO.

In 2010, Dehao Chen et al. \cite{Chen2010} continued the work started in Open64 and adapted it for GCC. In this work, several optimizations of GCC were specially adapted to the use of sampling profiles. The basic block and edges frequencies are derived using a Minimum Control Flow algorithm. In this solution, the Last Branch Record (LBR) precise sampling feature of the processor was used to improve the accuracy of the profile. Moreover, they also used a special version of the Lightweight Interprocedural Optimizer (LIPO) of GCC. The value profile is also derived from the sample profile using PEBS mode. With all these optimizations put together, they were able to achieve an average of 92\% of the performance gains of instrumentation-based Feedback Directed Optimization (FDO).

More recently, Dehao Chen (Google) released AutoFDO\footnote{\url{http://gcc.gnu.org/ml/gcc-patches/2012-09/msg01941.html}} (AFDO), on which our solution is based. It is a patch for GCC to handle sampling-based profiles. The profile is represented by a GCOV file, containing function profiles. Several optimizations of GCC have been reviewed to handle more accurately this new kind of profile. The profile is generated from the debug information contained into the executable and the samples are collected using the perf tool. AutoFDO is especially made to support optimized binary. For the time being, AFDO does not handle value profiles. Only the GCC patch has been released so far, no tool to generate the profile has been released.

More on the side of Performance Counters, Vincent M. Weaver shown that, when the setup is correctly tuned, the values of the performance counters have a very small variation between different runs (0.002 percent on the SPEC benchmarks). Nonetheless, very subtle changes in the setup can result in large variations in the results\cite{Weaver2008}.

Other sampling approaches without using performance counters have been proposed. For instance, The Morph system use statistical sampling of the program counter to collect profiles\cite{Zhang1997}. In another solution, kernel instructions were used to sample the contents of the branch-prediction unit of the hardware\cite{Conte1996}. These two solutions requires that additional information be encoded into the binary to correlates samples to the compiler's Intermediate Representation.

Performance Counters also start to be used in other areas than Profile-Guided Optimization. For instance, Schneider et al. sample performance counters to optimize data locality in VM with garbage collector\cite{Schneider2007}. In this solution, the collected data were used to driven online optimizations in the Just-In-Time (JIT) compiler of the VM.

\section{Conclusion and Future Work}
\label{sec:conclusion}

We designed and implemented a toolchain to use Hardware Event sampling to drive Profile-Guided Optimization inside GCC. Our implementation proved to be competitive with instrumentation-based PGO in terms of performance, achieving 93\% of the gains of traditional PGO and in terms of speed, having 15 times less overhead. The experiments show that this technique is already usable in production. However, its high performance is currently limited to C++ programs. Some work would have to be achieved to extend the current toolchain to support more programming languages. For that, the most important changes will need to be done in AFDO.

Instrumentation-based PGO has still an advantage over our implementation. It can generate value profiles. This kind of features is not yet supported by our toolchain. However, it has already been implemented with sampling-based PGO in \cite{Chen2010} and it something that was currently being developed in AFDO during our project, it should be integrated in the toolchain itself.

The presented toolchain makes it easy to handle new events. These events may lead to implementation of novel optimizations. Of course, sampling more events also incurs more overhead during profiling. Experiments have been made to integrate Load Latency events into GCC. The problem being that the new information is hard to use into existing or new optimization techniques. We implemented a Loop Fusion pass for GCC taking the Load Latency into account in its decision heuristics\cite{Wicht2013}. The main difficulty with Load Latency events being that they are not accurate enough at basic block level. To make better use of these events, it would be necessary to have a profile on an instruction level inside the compiler.

Some work will also have to be done to improve the speed of Gooda for large profiles, that is currently too slow.

\section{Acknowledgments}

We want to thank the reviewers for their comments and corrections regarding this paper. We would like to thank Stephane Eranian for his help in using the perf profiler and for providing kernel patches for perf events and libpfm.

\appendix
\section{Implementation}

The implementation of the profile generator is available on Github (\url{https://github.com/wichtounet/gooda-to-afdo-converter}).

The usage of the toolchain is described on the repository home page.

\bibliographystyle{plain}
\bibliography{bibliography}

\end{document}

%% file: template/preamble.tex
\usepackage{hyperref}
\usepackage{subcaption}

\usepackage[top=1in, bottom=1in, left=0.7in, right=0.7in]{geometry}

\usepackage{pgfplots}
\usepgfplotslibrary[units]
\pgfplotsset{compat=newest, width=8.5cm}

\hypersetup{
    colorlinks=true,        % false: boxed links; true: colored links
    linkcolor=black,        % color of internal links
    citecolor=black,        % color of links to bibliography
    filecolor=black,      % color of file links
    urlcolor=black           % color of external links
}